\documentclass[conference]{IEEEtran}

\usepackage{cite}
\usepackage{amsmath,amssymb,amsfonts}
\usepackage{graphicx}
\usepackage{textcomp}
\usepackage{xcolor}
\def\BibTeX{{\rm B\kern-.05em{\sc i\kern-.025em b}\kern-.08em
    T\kern-.1667em\lower.7ex\hbox{E}\kern-.125emX}}
\usepackage{url}
\usepackage[noend]{algorithmic}
\usepackage{xspace}
\usepackage{tabularx}
\usepackage{caption}
\usepackage{subcaption}
\usepackage[most]{tcolorbox}
\usepackage{enumitem}
\usepackage{xcolor}
\usepackage{amsfonts}
\usepackage{graphicx}
\usepackage{booktabs}
\usepackage[hidelinks]{hyperref}

\newcommand{\sysname}{TRACE\xspace}

\begin{document}

\title{\sysname: Topological Representations for Anomaly detection through Contrastive Embeddings}

\author{
\IEEEauthorblockN{Sydney Johns}
\IEEEauthorblockA{
Virginia Polytechnic Institute and State University\\
Alexandria, VA, USA\\
sydneyjohns@vt.edu
}
}

\maketitle

\begin{abstract}
Intrusion detectors must recognize evolving malicious behavior even when confirmed attack labels are scarce or incomplete. We present \sysname, a topology-aware graph contrastive learning framework for label-efficient intrusion detection. \sysname learns from benign system-activity graphs by combining message-passing and topological representations, then assigns anomaly scores in the same projected embedding space optimized by the contrastive objective. This design captures both local interactions and graph-scale structural deviations without using attack labels during representation learning. A labeled validation set containing benign and malicious
graphs is used only to calibrate the anomaly threshold. We evaluate \sysname on StreamSpot, GraSec-IoT, and Unicorn Wget against a supervised graph neural network and the benign-only GraphCL baseline. \sysname obtains F1 scores of 0.923, 0.790, and 0.583, respectively. Compared with GraphCL, this is higher on StreamSpot (0.450) and Unicorn Wget (0.203), but lower on GraSec-IoT (0.975). The supervised model leads on StreamSpot and GraSec-IoT, whereas \sysname has the highest F1 on Unicorn Wget. The mixed results show that topology-aware benign-only learning can improve detection in some graph settings but does not provide a uniform advantage.
\end{abstract}


\section{Introduction}\label{sec:introduction}

Modern IoT and enterprise systems generate large volumes of interactions among devices, hosts, services, processes, and files. Detecting intrusions in these environments is difficult because malicious behavior is rare, evolves over time, and is only partially represented in labeled datasets. Signature-based systems depend on known attack patterns, while conventional statistical detectors often rely on fixed assumptions about normal and abnormal behavior. Both can struggle when behavior changes or an attack does not match a predefined pattern \cite{Buczak2016,sommer2010outside}. Machine learning (ML)-based intrusion detection is attractive because it can learn behavioral representations from data and use them to identify suspicious deviations \cite{Buczak2016}.

The same adaptability, however, is limited when ML models require comprehensive attack labels. Confirmed malicious samples are expensive to obtain, attacks are much rarer than normal activity, and previously collected labels can become less representative as attacker behavior changes \cite{sommer2010outside,caville2022anomale}. A supervised detector may therefore learn a strong boundary for known attacks yet fail when deployment behavior differs from its labeled training set \cite{sommer2010outside}. We consequently study a benign-only representation-learning setting: the detector learns the distribution of normal behavior and flags deviations as potential intrusions \cite{manzoor2016streamspot,caville2022anomale}. This setting does not guarantee detection of every unseen attack, but it reduces dependence on labeled attacks and is relevant to resilient detection under incomplete or changing threat knowledge.

System activity is inherently relational, which makes graphs a natural representation for intrusion detection. A provenance graph records how system entities and events depend on or interact with one another. Its nodes can represent hosts, devices, services, processes, or files, and its edges can represent communication, access, or system events. A single event may appear harmless in isolation while its role in a larger interaction pattern reveals malicious behavior. For example, port scanning can generate one-to-many connections \cite{lee2003detection}, lateral movement can create multi-step paths across hosts \cite{fang2022lmtracker,zhou2024lmdetect}, and coordinated attacks can concentrate activity around selected targets \cite{bakar2024ftgnet}. Graph-based detection can therefore evaluate events in context rather than treating each record independently \cite{manzoor2016streamspot,lo2022egraphsage}.

This relational context makes topology important because topology describes the pattern of connectivity and graph shape. Attacks that unfold through a system can leave paths, branches, mergers of connected components, or changes in connectivity across processes, files, services, and hosts \cite{fang2022lmtracker,zhou2024lmdetect}. Such signatures summarize how entities are connected rather than only which entities or events occur. They may therefore distinguish graphs that have similar local attributes but different global organization. A topology-aware representation complements local message passing by preserving structural information across multiple scales, which is useful when deployment-time anomalies are expressed as changes in interaction structure.

Graph contrastive learning provides a way to learn these representations without attack labels. It creates augmented views of the same graph and trains a graph neural network (GNN) encoder to recognize that the views originate from the same graph. GraphCL demonstrated that this strategy can learn graph-level representations without supervised labels \cite{you2020graphcl}. Yet standard GNN message passing primarily aggregates local neighborhoods. Benign and malicious graphs can share common local patterns while differing in security-relevant paths, branches, or connectivity. TopoGCL addresses this limitation by adding summaries of graph shape across multiple scales, making topology-aware contrastive learning a suitable starting point for graph intrusion detection \cite{chen2024topogcl}.

Despite this promise, TopoGCL was designed mainly for unsupervised graph classification, not for anomaly scoring in an intrusion detection system \cite{chen2024topogcl}. Classification and intrusion detection have different goals. In graph classification, labeled examples can help the model learn class boundaries after representation learning. In intrusion detection, there are no labeled attack examples during representation learning. The detector must decide whether a new graph is abnormal by comparing it with the learned distribution. Because of this, the embedding space used for distance-based scoring directly controls detection quality.

To address this gap, we propose \sysname, a topology-aware graph contrastive learning method for intrusion detection. \sysname learns from augmented views of benign graphs, combines message-passing and topology-aware embeddings, and scores each test graph by its distance from benign training embeddings. Crucially, scoring occurs in the projected space optimized during contrastive training, avoiding a mismatch between the learned objective and the representation used for detection. The method requires neither attack labels during representation learning nor an additional classifier.

We evaluate \sysname on StreamSpot, GraSec-IoT, and Unicorn Wget \cite{manzoor2016streamspot,hamouche2024graseciot,unicornwget}. StreamSpot contains heterogeneous system-activity graphs. GraSec-IoT contains communication and attack behavior from an Internet of Things setting. Unicorn Wget contains graph samples from a system-activity attack scenario. Together, these datasets test whether \sysname works across different graph construction settings and attack behaviors. We compare \sysname with a supervised GNN and GraphCL, a graph contrastive learning baseline.

In summary, this paper makes the following contributions:
\begin{itemize}
    \item We formulate graph intrusion detection as benign-only representation learning and anomaly scoring under limited attack-label availability.
    \item We introduce \sysname, which combines message-passing and topological embeddings and performs detection in the contrastive space optimized during training.
    \item We evaluate \sysname on three graph security datasets against supervised and benign-only baselines, finding higher F1 than GraphCL on StreamSpot and Unicorn Wget but lower F1 on GraSec-IoT.
\end{itemize}

Together, these contributions examine how topology-aware, benign-only anomaly detection can support resilient ML-driven security monitoring when labeled attacks are incomplete or may not represent deployment behavior. The implementation and experimental results are available at \url{https://github.com/syd-kiwi/IDS_TopoGCL}.

\section{Related Work}\label{sec:related-work}

Graph neural networks can support intrusion detection by modeling relationships among network and system entities. Pujol-Perich et al.~\cite{pujolperich2021robustgnnids} show that graph structure can make network intrusion detection more robust than models that classify flows independently. E-GraphSAGE~\cite{lo2022egraphsage} applies a GNN to flow-based intrusion detection in Internet of Things environments and uses both edge features and graph structure. Recent surveys show that graph-based intrusion detection is a growing area, but that deployment, labeling, and robustness remain open challenges~\cite{GNN_IDS}. These works support the use of graphs for security monitoring. However, many graph-based intrusion detectors are supervised and depend on labeled non-malicious and malicious examples. In contrast, \sysname learns representations from non-malicious graphs.

Recent ARTMAN work places these technical choices in a broader trustworthy-ML context. Jung et al.~\cite{jung2023modelselection} address continuous operation under concept drift by selecting among models using a small amount of newly labeled data. Kolicic et al.~\cite{kolicic2024interpretable} emphasize interpretability and uncertainty awareness for online cybersecurity learning, while Miyamoto et al.~\cite{miyamoto2024featurefree} study feature-free Transformer-based traffic classification. More recently, Kawanaka et al.~\cite{kawanaka2025explainable} examine whether an ML-based network intrusion detector bases its decisions on attack-relevant characteristics. Together, these studies highlight adaptation, limited labeling, and analyst-facing trust as distinct deployment requirements. \sysname addresses the complementary problem of benign-only graph representation learning; it does not provide online adaptation, calibrated uncertainty, or localized explanations.

Provenance and temporal graph methods are also closely related to our task. These methods use graph structure to describe how system activity changes over time. Jbeil~\cite{Jbeil} studies temporal graph learning for lateral movement detection in enterprise networks. MAGIC~\cite{MAGIC} uses masked graph representation learning to detect advanced persistent threats from provenance graphs. FLASH~\cite{Flash} and KAIROS~\cite{cheng2024kairos} also use provenance graphs to support intrusion detection and investigation. These systems show that graph-structured representations are useful for security monitoring. Our work differs in focus. We study graph-level anomaly scoring for benign-only representation learning and focus on how topology-aware contrastive embeddings should be used for detection.

Graph contrastive learning is significant for this problem because it can learn useful graph embeddings without attack labels. GraphCL~\cite{you2020graphcl} learns graph-level representations by creating augmented views of the same graph and training an encoder to make their embeddings agree. This makes GraphCL an important baseline for our work. However, GraphCL does not explicitly model higher-order graph shape. This can be a limitation when non-malicious and malicious activity graphs share the same broad structure but differ in smaller structural details. Graph-JEPA offers another self-supervised approach to graph-level representation learning by predicting masked subgraph representations from a context subgraph, without contrastive negative pairs \cite{skenderi2025graphjepa}. TRACE instead uses a topology-aware contrastive objective and nearest-neighbor anomaly scoring for intrusion detection. Evaluating predictive graph representations under the same benign-only learning protocol is a direction for future work.

TopoGCL~\cite{chen2024topogcl} addresses this limitation by adding topological information to graph contrastive learning. It uses topological summaries to describe graph shape across multiple scales. This makes TopoGCL the closest methodological starting point to \sysname. However, TopoGCL was designed mainly for unsupervised graph classification, not anomaly scoring or intrusion detection. A graph classifier can use labeled examples after representation learning to separate classes. A model trained only on non-malicious graphs cannot rely on labeled attacks during representation learning. It must decide whether a test graph is abnormal by comparing it with non-malicious training graphs.

\sysname builds on graph contrastive learning and topology-aware graph learning, but it targets a different security setting. It adapts topology-aware contrastive learning for intrusion detection and scores test graphs by their distance from non-malicious training embeddings. The key difference is that \sysname scores graphs in the same projected embedding space shaped by the contrastive loss. This keeps representation learning and anomaly scoring consistent. The goal is not to introduce a new topological summary. Instead, \sysname uses topology-aware contrastive embeddings more directly for intrusion detection in intelligent systems.

\section{Methodology}\label{sec:method}

\begin{figure*}[t]
    \centering
    \includegraphics[width=\textwidth]{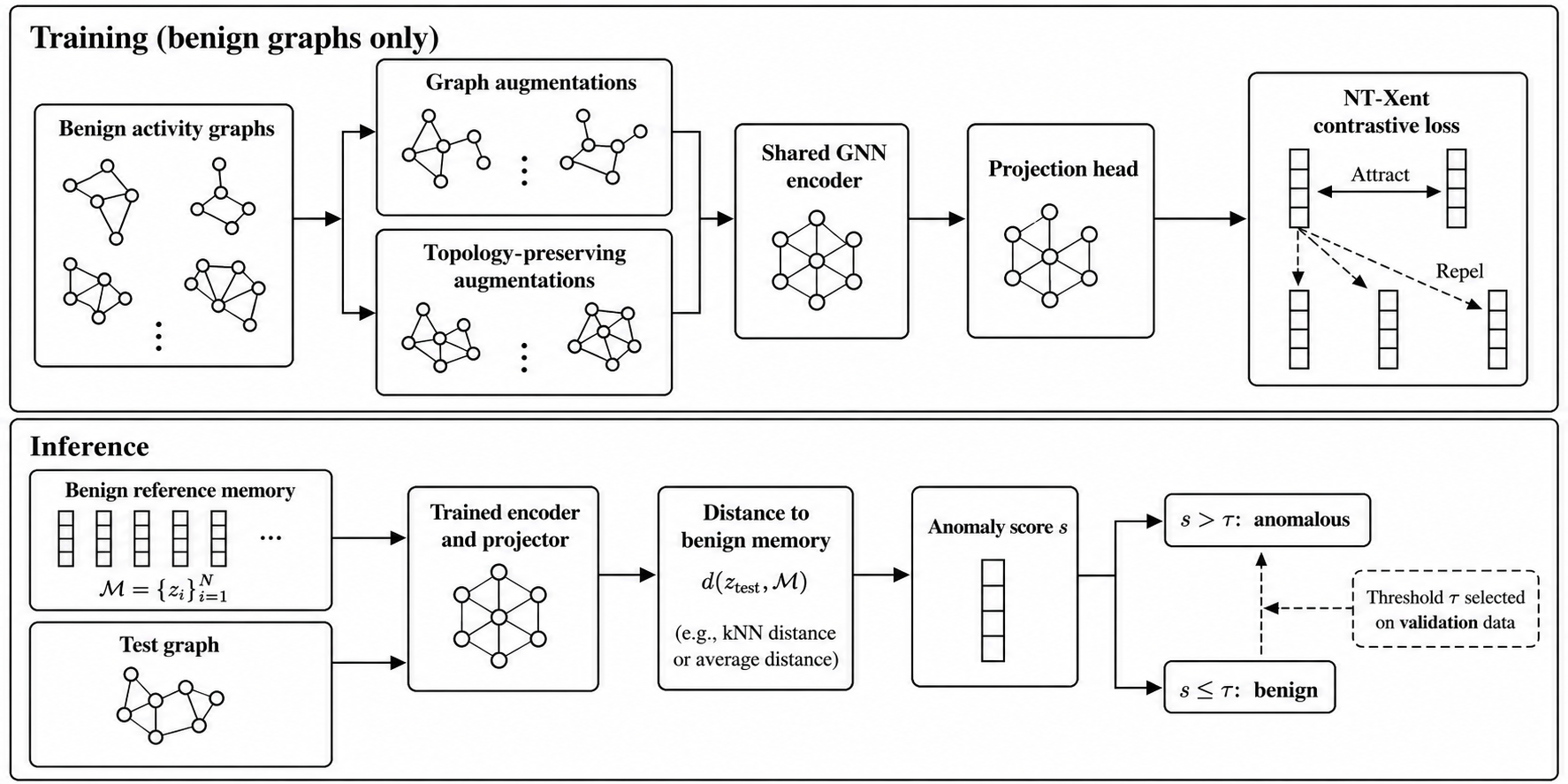}
    \caption{
    Overview of \sysname{}. The top row shows contrastive training. Activity graphs are augmented into graph and topology views, encoded, projected, and optimized with a contrastive objective. The bottom row shows inference. Each graph is scored by distance to the non-malicious reference memory and classified using a validation-selected threshold.
    }
    \label{fig:method}
\end{figure*}

\sysname{} is an anomaly detection method for graph-based intrusion detection. It has four main steps. First, system activity is represented as a graph. Second, \sysname{} learns graph and topology embeddings from non-malicious training graphs using contrastive learning. Third, each validation or test graph is embedded in the same contrastive space used during training. Fourth, the detector assigns an anomaly score by measuring distance from non-malicious training embeddings. The subsections below follow this pipeline: problem setup, graph representation, contrastive training, anomaly scoring, and thresholding.

Let $G_i=(V_i,E_i,X_i)$ denote the $i$th activity graph. Here, $V_i$ is the set of nodes, $E_i$ is the set of edges, and $X_i$ is the node feature matrix. Nodes can represent entities such as hosts, devices, services, processes, or files. Edges represent interactions such as communication, access, or system events.

Let $\mathcal{B}_{\text{train}}=\{G_i : y_i=0\}$ be the training set, where $y_i=0$ means non-malicious and $y_i=1$ means malicious. During representation learning, \sysname{} uses only graphs from $\mathcal{B}_{\text{train}}$. Malicious graphs are not used to train the encoder or projection head. They are used only during validation for threshold selection and during testing for evaluation.

The goal is to learn an embedding function $f(\cdot)$ that maps each graph $G_i$ to a vector representation. At test time, graphs that are far from the non-malicious training distribution receive higher anomaly scores. This setting differs from supervised graph classification. A supervised graph neural network learns directly from both benign and malicious labels. \sysname{} instead learns normal behavior from benign graphs and detects graphs that do not match that learned normal pattern.

\subsection{Operational Setting and Threat Model}

\sysname{} is intended for monitoring environments in which system or network telemetry can be grouped into activity graphs but comprehensive attack labels are unavailable. Provenance- and communication-graph systems demonstrate how interactions can support intrusion detection and investigation \cite{manzoor2016streamspot,cheng2024kairos,lo2022egraphsage}. A graph summarizes the entities and interactions observed within the unit of analysis defined by a dataset, such as a system-activity trace or an IoT communication snapshot. The detector receives one complete graph at a time and returns a graph-level anomaly score and binary decision. In an operational workflow, this decision would identify activity graphs for further investigation; it does not by itself identify the responsible process, connection, or subgraph. The construction and duration of each graph therefore determine the granularity of both detection and response.

The learning setting separates representation learning from threshold calibration. The encoder and projection heads learn exclusively from graphs designated as benign, so their parameters are not optimized with attack-class labels. This assumption reflects the difficulty of obtaining representative attack labels and transferring closed-world ML evaluations to operational intrusion detection \cite{sommer2010outside}. It also avoids requiring the training set to enumerate the attack behaviors that may later appear, consistent with prior self-supervised graph anomaly detection motivated by limited labels \cite{caville2022anomale}. Nevertheless, the evaluation is not completely label-free: the operating threshold is selected with a validation set containing benign and malicious graphs. Thus, ``benign-only'' in this paper refers specifically to representation learning. Deployment without labeled attacks would require another calibration strategy, such as a quantile of benign validation scores, and that alternative is outside the present experimental study.

The central detection assumption is that malicious activity produces a measurable departure from the learned benign embedding distribution, as in graph-stream anomaly detection \cite{manzoor2016streamspot}. \sysname{} is particularly motivated by departures expressed through graph structure: new paths, branches, changes in connectivity, or altered relationships among entities. The graph channel represents attributed local neighborhoods and their pooled graph-level summary, while the topology channel preserves complementary information about connectivity across a filtration, following the motivation for topology-aware graph contrastive learning \cite{chen2024topogcl}. Their combination is designed to make structurally unusual activity distant from nearby benign examples. The method does not assume that a test graph belongs to a previously observed attack class, but it cannot guarantee detection of an unseen attack. An attack that closely reproduces benign attributes and topology may receive a low anomaly score.

The benign reference set is also a trust assumption. Substantial contamination by undetected attacks could cause malicious behavior to become part of the learned normal distribution, reducing its anomaly score at inference. Likewise, legitimate but previously unseen operational changes can be distant from the reference set and produce false alerts. These two failure modes reflect broader deployment and robustness challenges identified for graph-based intrusion detection \cite{GNN_IDS}. They create a practical tension: a reference memory that is too static may not represent benign behavioral drift, whereas indiscriminate updating could absorb malicious or unverified behavior. The current study uses fixed training, validation, and test splits and does not implement online adaptation. Safe reference-memory updates, drift detection, and contamination-resistant learning remain deployment requirements rather than evaluated capabilities.

Within this scope, resilience means reducing dependence on a complete and permanently representative catalog of labeled attacks. \sysname{} supports this objective by learning a benign structural reference and treating deviations as candidates for investigation. Its topology-aware representation may remain informative when local event attributes vary but the organization of system interactions changes. At the same time, trustworthy use requires presenting the anomaly score as evidence for triage rather than proof of compromise. Thresholds must reflect the relative costs of missed attacks and false alerts, and analysts need access to the underlying telemetry before acting on a prediction.

This threat model does not include an attacker who deliberately manipulates training data, graph construction, features, or augmentations to evade the detector. Structural adversarial attacks against graph-based intrusion detectors constitute a distinct evaluation setting \cite{venturi2024structuralattacks}. The experiments also do not evaluate poisoning, adaptive evasion, compromised telemetry, or certified robustness. Accordingly, our use of resilience concerns detection under limited labels and differences between learned benign behavior and deployment observations; it is not a claim of adversarial robustness. Explicit evaluation against adaptive attackers is an important direction for future work.

\subsection{Graph and Topology Representations}

\sysname{} uses two complementary views of each graph: a graph-structure view and a topology view.

The graph-structure view is learned with a graph neural network encoder. In our implementation, this encoder is a graph isomorphism network. The encoder performs message passing over the graph and pools node embeddings into one graph-level readout. We use sum, mean, and max pooling and concatenate the three pooled vectors. This produces a raw graph readout, denoted by $h_i$.

The raw readout $h_i$ is then passed through a projection head. The projection head maps $h_i$ into a contrastive graph embedding, denoted by $z_i^{G}$. This embedding is the graph-channel representation used by the contrastive loss.

The topology view captures graph shape more directly. For each graph, \sysname{} computes topological summaries from structural node scores such as degree and closeness. These summaries describe how connected components appear and merge as the graph is viewed through a filtration. The topological summaries are converted into fixed-length vectors and passed through a small multilayer perceptron. The output is a topology embedding, denoted by $z_i^{T}$.

The final contrastive embedding for graph $G_i$ combines the graph-channel and topology-channel embeddings:
\begin{equation}
    e_i =
    \left[
    \operatorname{norm}(z_i^{G})
    \; \Vert \;
    \operatorname{norm}(z_i^{T})
    \right],
    \label{eq:joint-embedding}
\end{equation}
where $\operatorname{norm}(\cdot)$ denotes $\ell_2$ normalization and $\Vert$ denotes vector concatenation. Eq.~\eqref{eq:joint-embedding} defines the embedding used later for anomaly scoring. If only one channel is used in an ablation, then $e_i$ contains only that channel.

\subsection{Contrastive Training}

During training, \sysname{} creates two augmented views of each graph. These views preserve the main identity of the graph while changing small details. For the graph channel, we use graph augmentations such as node dropping, edge dropping, and feature masking. For the topology channel, we perturb the topology feature vector with masking and noise.

InfoNCE trains the representation by posing a matching problem within each batch. For every embedding from the first augmented view, the corresponding embedding of the same original graph in the second view is the positive match; embeddings of the other graphs are alternatives. The loss increases the positive pair's similarity relative to all alternatives, with the temperature controlling how strongly it emphasizes the closest matches. Thus, the objective preserves graph identity under small augmentations while separating different benign activity patterns. Formally, for a batch of $N$ graphs, let $A=\{a_1,\ldots,a_N\}$ and $B=\{b_1,\ldots,b_N\}$ denote the embeddings of the two views. The one-direction InfoNCE loss is:
\begin{equation}
    \mathcal{L}_{\text{NCE}}(A,B)
    =
    -\frac{1}{N}
    \sum_{i=1}^{N}
    \log
    \frac{
        \exp(\operatorname{sim}(a_i,b_i)/\tau)
    }{
        \sum_{j=1}^{N}
        \exp(\operatorname{sim}(a_i,b_j)/\tau)
    },
    \label{eq:infonce}
\end{equation}
where $\operatorname{sim}(\cdot,\cdot)$ is cosine similarity and $\tau$ is a temperature parameter. Eq.~\eqref{eq:infonce} makes embeddings from two views of the same graph closer and pushes embeddings from different graphs apart. In practice, we use the symmetric version by averaging $\mathcal{L}_{\text{NCE}}(A,B)$ and $\mathcal{L}_{\text{NCE}}(B,A)$.

The full training loss combines the graph-channel contrastive loss and the topology-channel contrastive loss:
\begin{equation}
    \mathcal{L}
    =
    \alpha \mathcal{L}_{G}
    +
    \beta \mathcal{L}_{T},
    \label{eq:training-loss}
\end{equation}
where $\mathcal{L}_{G}$ is the graph-channel loss, $\mathcal{L}_{T}$ is the topology-channel loss, and $\alpha$ and $\beta$ control their weights. Eq.~\eqref{eq:training-loss} trains \sysname{} to learn both local graph structure and higher-level topology from graphs.

\subsection{Contrastive-Space Anomaly Scoring}

After training, \sysname{} embeds every training graph using the trained model. Let
$\mathcal{E}_{\text{benign}}=\{e_j : G_j \in \mathcal{B}_{\text{train}}\}$
be the set of benign training embeddings.

For a validation or test graph $G_i$, \sysname{} computes its embedding $e_i$ using Eq.~\eqref{eq:joint-embedding}. It then compares $e_i$ with the training embeddings. The anomaly score is the average distance to the $k$ nearest embeddings:
\begin{equation}
    s(G_i)
    =
    \frac{1}{k}
    \sum_{e_j \in \operatorname{kNN}(e_i,\mathcal{E}_{\text{benign}})}
    \lVert e_i-e_j\rVert_2 .
    \label{eq:knn-score}
\end{equation}
Eq.~\eqref{eq:knn-score} assigns a larger score to graphs that are farther from normal training graphs. These graphs are more likely to be anomalous.

The key design choice in \sysname{} is that anomaly scoring uses the same contrastive embedding space shaped during training. Some contrastive pipelines train a projection head but later score graphs using the raw encoder readout $h_i$. This can create a mismatch because the contrastive objective directly optimizes the projected embedding, not the raw readout. \sysname{} avoids this mismatch by scoring graphs with $e_i$, which is built from the contrastive embeddings. This keeps training and detection consistent.

\subsection{Detection and Thresholding}

The anomaly score $s(G_i)$ is converted into a binary prediction using a threshold $\gamma$:
\begin{equation}
    \hat{y}_i =
    \begin{cases}
    1, & s(G_i) \geq \gamma,\\
    0, & s(G_i) < \gamma,
    \end{cases}
    \label{eq:threshold}
\end{equation}
where $\hat{y}_i=1$ means the graph is predicted as malicious and $\hat{y}_i=0$ means it is predicted as non-malicious. Eq.~\eqref{eq:threshold} defines the final detection rule.

In our experiments, $\gamma$ is selected on the validation set. The validation set contains both non-malicious and malicious graphs so that we can choose the threshold that gives the best validation F1 score. The selected threshold is then fixed and applied once to the test set. The encoder and projection head are still trained only on benign graphs. Malicious labels are used for threshold selection and evaluation, not for representation learning.

\section{Evaluation}\label{sec:evaluation}

\begin{table*}[t]
\centering
\caption{Intrusion detection performance across three graph-based security datasets. The supervised GNN uses labeled attacks during training. GraphCL and \sysname{} are benign-only representation learning methods.}
\label{tab:main_results}
\resizebox{\textwidth}{!}{%
\begin{tabular}{llcccc}
\toprule
Dataset & Model & Accuracy & Precision & Recall & F1 \\
\midrule
StreamSpot & Supervised GNN & $0.989 \pm 0.016$ & $1.000 \pm 0.000$ & $0.933 \pm 0.094$ & $0.963 \pm 0.052$ \\
StreamSpot & GraphCL & $0.750 \pm 0.036$ & $0.353 \pm 0.054$ & $0.733 \pm 0.377$ & $0.450 \pm 0.154$ \\
StreamSpot & \sysname{} & $0.917 \pm 0.023$ & $0.862 \pm 0.037$ & $0.996 \pm 0.006$ & $0.923 \pm 0.019$ \\
\midrule
GraSec-IoT & Supervised GNN & $0.988 \pm 0.003$ & $0.997 \pm 0.002$ & $0.990 \pm 0.004$ & $0.994 \pm 0.002$ \\
GraSec-IoT & GraphCL & $0.953 \pm 0.006$ & $0.970 \pm 0.006$ & $0.980 \pm 0.008$ & $0.975 \pm 0.003$ \\
GraSec-IoT & \sysname{} & $0.742 \pm 0.124$ & $0.718 \pm 0.129$ & $0.909 \pm 0.064$ & $0.790 \pm 0.065$ \\
\midrule
Unicorn Wget & Supervised GNN & $0.867 \pm 0.060$ & $0.600 \pm 0.490$ & $0.267 \pm 0.226$ & $0.367 \pm 0.306$ \\
Unicorn Wget & GraphCL & $0.560 \pm 0.200$ & $0.123 \pm 0.101$ & $0.600 \pm 0.490$ & $0.203 \pm 0.167$ \\
Unicorn Wget & \sysname{} & $0.515 \pm 0.062$ & $0.537 \pm 0.071$ & $0.710 \pm 0.215$ & $0.583 \pm 0.053$ \\
\bottomrule
\end{tabular}%
}
\end{table*}

This section evaluates whether \sysname{} provides effective graph-level intrusion detection across datasets, baselines, and label-availability settings.

\subsection{Research Questions}

We evaluate three research questions:

\begin{itemize}
    \item RQ1: How does \sysname{} compare with supervised and benign-only graph baselines across intrusion detection datasets?

    \item RQ2: Does topology-aware contrastive scoring improve over standard graph contrastive learning?

    \item RQ3: How do the methods balance attack recall and alert precision?
\end{itemize}

\subsection{Datasets}

We evaluate graph-level intrusion detection on three datasets. Each dataset is converted into a set of graphs. Each graph has a benign or malicious label for validation and testing. During representation learning, benign-only methods use only benign training graphs.

\begin{itemize}
    \item \textbf{StreamSpot:} contains heterogeneous system-activity provenance graphs. These graphs represent relationships among system entities and events. StreamSpot is useful for testing whether a detector can identify malicious behavior in structured system activity \cite{manzoor2016streamspot}.

    \item \textbf{GraSec-IoT:} contains graph snapshots from an Internet of Things security setting. It includes communication behavior and attack activity in an IoT environment \cite{hamouche2024graseciot}. This dataset tests whether the method works on IoT-style communication graphs.

    \item \textbf{Unicorn Wget:} is a graph-based security dataset built from system activity collected from a Wget attack scenario. It provides graph samples rather than tabular flow records, so it fits our graph-level detection setting \cite{unicornwget}.
\end{itemize}

The datasets contain 600 StreamSpot graphs, 7,443 GraSec-IoT graphs, and 150 Unicorn Wget graphs. For TRACE, StreamSpot uses 360 benign training graphs, 40 validation graphs, and 160 test graphs. GraSec-IoT uses 342 benign training graphs, 152 validation graphs, and 44 test graphs. Unicorn Wget uses 90 benign training graphs, 10 validation graphs, and
40 test graphs. These split sizes remain constant across the reported
random seeds. The methods use the same reported random seeds but different data partitions in the saved experiments. Consequently, the results compare the evaluated configurations rather than performance on identical graphs.

\subsection{Baselines}

We compare \sysname{} with one label-rich reference model and one benign-only representation learning baseline.

\begin{itemize}
    \item \textbf{Supervised GNN.} This supervised graph neural network is trained with labeled benign and malicious graphs. We treat it as a label-rich reference baseline. It is not directly comparable to benign-only representation learning methods because it uses attack labels during training, but it shows what performance is possible when labels are available.

    \item \textbf{GraphCL.} This benign-only graph contrastive learning baseline learns graph embeddings from augmented graph views without using attack labels during representation learning \cite{you2020graphcl}. It tests whether standard graph contrastive learning is sufficient for this task.
\end{itemize}

\subsection{Experimental Setup}

All methods use the same split logic for a given dataset and seed. StreamSpot and GraSec-IoT results aggregate seeds 42, 43, and 44; Unicorn Wget results aggregate seeds 42 through 46. The threshold is selected separately on the validation set for each benign-only run before being applied once to the corresponding test set.

The encoder is trained only on benign training graphs. Malicious graphs are used only for validation threshold selection and final testing. This prevents attack labels from being used during representation learning.

For \sysname{}, we use 12 training epochs, hidden dimension 64, embedding dimension 64, temperature 0.2, maximum 512 nodes per graph, node drop rate 0.10, edge drop rate 0.05, feature mask rate 0.10, topology mask rate 0.10, topology noise 0.10, and $k=5$ nearest neighbors for anomaly scoring. For the joint graph-topology model, we use $\alpha=0.5$ and $\beta=0.5$ to weight the graph and topology contrastive losses. Baseline hyperparameters follow their saved experimental configurations rather than being assumed identical to \sysname{}.

\subsection{Metrics}

We report accuracy, precision, recall, and F1 score. Let TP be true positives, TN be true negatives, FP be false positives, and FN be false negatives. Precision is $\frac{TP}{TP+FP}$ and measures how many predicted attacks are actually attacks. Recall is $\frac{TP}{TP+FN}$ and measures how many attacks are detected. F1 score is the harmonic mean of precision and recall.

F1 is useful because intrusion detection needs both high precision and high recall. Precision is important because too many false alerts can make an intrusion detection system difficult to use in practice. Recall is important because missed attacks can leave a system exposed.

\subsection{Results by Research Question}

Table~\ref{tab:main_results} reports intrusion detection performance across StreamSpot, GraSec-IoT, and Unicorn Wget. The supervised GNN uses labeled benign and malicious graphs during training. GraphCL and \sysname{} train their encoders using only benign graphs and use labeled validation data only for threshold selection.

\noindent\textbf{RQ1: Comparison with supervised and benign-only baselines.}
The supervised GNN achieves the highest F1 on StreamSpot ($0.963 \pm 0.052$) and GraSec-IoT ($0.994 \pm 0.002$), while \sysname{} achieves the highest F1 on Unicorn Wget ($0.583 \pm 0.053$ versus $0.367 \pm 0.306$ for the supervised GNN). Among the benign-only methods, \sysname{} has higher F1 on StreamSpot ($0.923 \pm 0.019$ versus $0.450 \pm 0.154$) and Unicorn Wget ($0.583 \pm 0.053$ versus $0.203 \pm 0.167$), whereas GraphCL leads on GraSec-IoT ($0.975 \pm 0.003$ versus $0.790 \pm 0.065$). Thus, \sysname{} is competitive in two benign-only comparisons but is neither uniformly superior to GraphCL nor a replacement for label-rich learning when representative attack labels are available.

\noindent\textbf{RQ2: Comparison with standard graph contrastive learning.}
Topology-aware contrastive scoring does not improve over standard GraphCL uniformly. \sysname{} has higher F1 by $0.473$ on StreamSpot and $0.380$ on Unicorn Wget, but lower F1 by $0.185$ on GraSec-IoT. This heterogeneity indicates that the value of degree- and closeness-based topological summaries depends on the dataset and graph construction. A plausible explanation is that topology adds complementary information when anomalous behavior changes connectivity at graph scale, while the standard graph representation is already sufficient for the GraSec-IoT split. Because the methods also differ in architecture and training configuration, these comparisons do not isolate topology as the sole causal factor.

\noindent\textbf{RQ3: Precision--recall balance.}
On StreamSpot, \sysname{} combines recall of $0.996 \pm 0.006$ with precision of $0.862 \pm 0.037$, whereas GraphCL has lower and much more variable recall ($0.733 \pm 0.377$) and precision ($0.353 \pm 0.054$). On GraSec-IoT, GraphCL provides the stronger balance, with precision of $0.970 \pm 0.006$ and recall of $0.980 \pm 0.008$; \sysname{} retains high recall ($0.909 \pm 0.064$) but lower precision ($0.718 \pm 0.129$). On Unicorn Wget, \sysname{} improves the benign-only balance, but both benign-only methods and the supervised GNN exhibit substantial weaknesses or variability. For resilient monitoring, the preferred operating point remains deployment-dependent: missed attacks favor recall, whereas analyst capacity and alert fatigue favor precision.

Overall, RQ1 shows that the supervised reference has the highest F1 on StreamSpot and GraSec-IoT, while \sysname{} leads on Unicorn Wget and is the stronger benign-only method on StreamSpot and Unicorn Wget. RQ2 shows that topology-aware scoring is dataset-dependent rather than uniformly beneficial. RQ3 shows a strong TRACE recall--precision balance on StreamSpot, a stronger GraphCL balance on GraSec-IoT, and challenging, variable behavior for all methods on Unicorn Wget.

\section{Discussion and Limitations}\label{sec:discussion}

The results answer the three research questions with a qualified conclusion. For RQ1, the supervised GNN has the highest F1 on StreamSpot and GraSec-IoT, while \sysname{} leads on Unicorn Wget; among the benign-only methods, \sysname{} is stronger on StreamSpot and Unicorn Wget and GraphCL is stronger on GraSec-IoT. For RQ2, topology-aware scoring is associated with higher F1 on two datasets and lower F1 on one, so the evaluation does not support a universal benefit. For RQ3, \sysname{} balances precision and recall well on StreamSpot, GraphCL does so on GraSec-IoT, and all methods show limitations or high variation on Unicorn Wget. These outcomes argue for dataset-dependent utility rather than overall superiority.

The supervised GNN's advantage on StreamSpot and GraSec-IoT is fair and expected because it is optimized directly with both benign and malicious labels, while \sysname{} and GraphCL do not use attack labels during representation learning. When labeled training attacks resemble test attacks, supervision provides direct information for locating a decision boundary. On Unicorn Wget, however, the supervised GNN's high accuracy is accompanied by low and variable recall and a lower F1 than \sysname{}. Accuracy alone therefore does not imply reliable attack coverage. The available results do not establish the cause; possible contributors include the smaller sample, graph construction, and the relationship between labeled training and test behavior. We avoid attributing the behavior to overfitting without further analysis. Matched ablations that remove the topology channel and compare scoring before and after projection are needed to establish the individual contributions of these choices.

Dataset construction provides a cautious explanation for the mixed topology-aware results. StreamSpot contains provenance graphs of heterogeneous system activity, GraSec-IoT represents IoT communication snapshots, and Unicorn Wget captures a specific system-activity scenario. These graph types can encode attacks at different structural scales. Degree- and closeness-based summaries may add complementary information for the two provenance-style settings, whereas GraphCL's learned representation is more effective on the evaluated GraSec-IoT split. Confirming this interpretation would require attack-level structural analysis and matched ablations beyond the present study.

From an ARTMAN perspective, \sysname{} addresses one aspect of resilient and trustworthy ML-driven security: maintaining an anomaly-detection capability when comprehensive attack labels are unavailable. Learning a benign reference and scoring deviations reduces reliance on a fixed catalog of attacks. Prior ARTMAN studies show that continuous security operation also requires mechanisms for concept drift, uncertainty, and explanation~\cite{jung2023modelselection,kolicic2024interpretable,kawanaka2025explainable}. The present evaluation does not establish those capabilities: its reference memory is fixed, its scores are not calibrated as predictive uncertainty, and its alerts are not localized. This is also not a claim of adversarial robustness. The evaluation measures ordinary detection performance under the reported splits; it does not test evasion, poisoning, adaptive attackers, or certified robustness.

\subsection{Limitations and Threats to Validity}

Several limitations constrain the conclusions. First, only three datasets and their existing graph constructions are evaluated, so the results may not generalize to other networks, organizations, attack families, or graph-generation procedures. Second, the comparison does not use a fully matched architecture and hyperparameter configuration across GraphCL and \sysname{}, so it cannot attribute performance differences solely to topology. Third, labeled malicious validation graphs are used to select the F1-maximizing threshold. Although attack labels do not train the representation, this choice is not fully label-free and may be difficult to reproduce when deployment-time attacks are unavailable.

Fourth, balanced evaluation sets and F1-selected thresholds do not reproduce the low attack prevalence of many operational environments. Precision and alert volume may change under realistic class imbalance. Fifth, the supervised and benign-only methods have different access to labels, so the supervised GNN is a label-rich reference rather than a strictly matched baseline. Sixth, graph truncation, augmentations, hyperparameters, and the selected degree and closeness filtrations may affect performance; the present experiments do not establish which design component causes each dataset-level difference. Finally, \sysname{} detects anomalies at graph level and does not localize the nodes, edges, or subgraphs responsible for an alert, limiting analyst-facing explanation.

Future work should evaluate additional attack families and deployment shifts, select thresholds without labeled attacks, test under realistic class imbalance, and study sensitivity to graph construction and topology choices. Attack localization and structural explanations would make alerts more actionable. Robustness to adversarial manipulation, poisoning, and adaptive evasion also requires separate experiments before \sysname{} can be characterized as adversarially robust.

\section{Conclusion}\label{sec:conclusion}

This paper studied topology-aware graph contrastive learning for benign-only graph-based intrusion detection. We introduced \sysname{}, a method that learns from benign activity graphs and scores test graphs by distance from benign training embeddings. \sysname{} combines graph-level message-passing representations with topology-aware structural information and performs anomaly scoring in the same contrastive embedding space used during training.

Across three graph-based security datasets, \sysname{} obtains F1 scores of $0.923$ on StreamSpot, $0.790$ on GraSec-IoT, and $0.583$ on Unicorn Wget. It outperforms GraphCL on StreamSpot and Unicorn Wget but underperforms it on GraSec-IoT. The supervised GNN leads on StreamSpot and GraSec-IoT, whereas \sysname{} has the highest F1 on Unicorn Wget. These mixed results show that topology-aware contrastive anomaly scoring can strengthen benign-only detection in some graph settings but is not uniformly beneficial. The result supports continued study of topology-aware detection under incomplete attack labels, while the limited datasets, unmatched baseline configurations, and label-dependent threshold selection motivate broader and more controlled validation.

\section*{LLM Usage Statement}
LLMs were used for editorial purposes in this manuscript, and all outputs were inspected by the authors to ensure accuracy and originality. Their use was limited to language editing and organization. The authors take responsibility for the manuscript’s correctness, originality, citations, and scientific claims. No experiments, experimental data, or numerical results were generated by these tools.

\bibliographystyle{ieeetr}
\bibliography{ref}

\end{document}